\documentclass[aps,prd,nofootinbib,twocolumn,showpacs]{revtex4-1}
\usepackage[english]{babel}
\usepackage[utf8]{inputenc}
\usepackage{amsmath,ulem}
\usepackage{amssymb}
\usepackage{graphicx}
\usepackage{color}
\usepackage{float}
\usepackage{slashed}
\usepackage{lipsum}

\usepackage{hyperref}

\newcommand{\ev}[1]{\ensuremath{\big{<}#1\big{>}}}
\newcommand\numberthis{\addtocounter{equation}{1}\tag{\theequation}}

\DeclareMathOperator{\diag}{diag}

\begin{document}
	
\title{Hybrid stars with large strange quark cores constrained by GW170817}

\author{Márcio Ferreira}
\email{marcio.ferreira@uc.pt}
\affiliation{CFisUC, 
	Department of Physics, University of Coimbra, P-3004 - 516  Coimbra, Portugal}
	
\author{Renan Câmara Pereira}
\email{renan.pereira@student.uc.pt}
\affiliation{CFisUC, Department of Physics, University of Coimbra, 
	P-3004 - 516  Coimbra, Portugal}

\author{Constança Providência}
\email{cp@fis.uc.pt}
\affiliation{CFisUC, 
	Department of Physics, University of Coimbra, P-3004 - 516  Coimbra, Portugal}

\date{\today}

\begin{abstract}
We analyze the impact of several multi-quark interaction channels on
the properties of hybrid stars. Four and eight vector interactions
were included in the $SU(3)$ NJL model to describe the quark matter,
and all vector like interactions were investigated. The hybrid stars were built starting from an hadronic EoS that satisfy presently accepted nuclear matter properties and considering the quark model 
constrained by the vacuum properties of several  light mesons. The
interplay between the 8-quark vector interaction and the 4-quark
isovector-vector interaction determine the size of the quark core and
its strangeness content. The inclusion of a
isovector-vector interaction is essential to
obtain large $s$-quark contents and a smaller central speed of sound, although this interaction pushes the
onset of quarks to larger densities. It is observed that low mass stars,
with a mass below 1.4$M_\odot$, may contain a quark core but with
small strangeness content. 
It is shown that the presently existing mass and tidal
deformability constraints from NS observations
allow for the existence of hybrid stars with a large strangeness
content and large quark cores for binary mass ratios $q=M_2/M_1\gtrsim 0.85$. 

\end{abstract}

\maketitle

\section{Introduction}
\label{introduction}

Multi-messenger astrophysics provides deeper insights into neutron stars (NS) physics by combining astrophysical observations of
electromagnetic radiation and gravitational waves (GW). 
The observation of the pulsars PSR J1614-2230 ($M=1.908\pm$0.016
$M_\odot$) \cite{Demorest2010,Arzoumanian2017}, PSR J0348+0432 ($M=2.01\pm$0.04 $M_\odot$)
\cite{Antoniadis:2013pzd}, and  MSP J0740+6620 \cite{Cromartie2019},
($M=2.14^{+0.10}_{-0.09} M_\odot$)  established considerable constraints on the nuclear matter equation of state (EoS).
The analysis by the LIGO/Virgo collaboration of the GW170817 NS merger 
provided important information about NS physics
\cite{TheLIGOScientific:2017qsa,PhysRevLett.121.161101}, e.g., an
upper limit of the tidal deformability of a NS star, constraining the high density EoS. 
The observations of the gamma-ray burst (GRB) GRB170817A \cite{grb} and 
electromagnetic transient AT2017gfo \cite{kilo}, which followed up the
GW170817 event, give further hints on the NS tidal deformability \cite{Radice2017,Radice2018,Bauswein:2019skm,Coughlin2018,Wang2018}.
A recent observation from the LIGO/Virgo collaboration is the compact binary coalescence GW190814 \cite{Abbott:2020khf}. Even though the primary 
binary component is conclusively a black hole of mass
$22.2-24.3M_{\odot}$, the nature of the secondary component with a mass
$2.50-2.67M_{\odot}$ remains yet inconclusive \cite{Abbott:2020khf}.
Recent estimates for both the mass and radius of the
millisecond-pulsar PSR J0030+0451 were put forward by the Neutron Star
Interior Composition Explorer (NICER) mission
\cite{Riley_2019}. However, the uncertainties obtained are still large
and therefore, do not constrain much the EoS.\\

The NS tidal deformability estimated from GW observations favors a not
too stiff EoS for NS matter \cite{Abbott18}. However, this behavior conflicts with the observations of massive pulsars, 
which require a stiff EoS and predict large radii and thus large tidal deformabilities.
The contrasting indications of low radii predictions from tidal deformabilities estimations and 
large radii from massive pulsar observations, pose considerable
constraints on the high density region of the nuclear matter EoS.
These indications might be a signal of the existence of non-nucleonic degrees of freedom at high densities in NS matter,
such as quark matter. The existence of a first-order phase transition from hadronic to quark matter may still explain the observational data \cite{Alford:2019oge}. Detecting observational signatures for the presence of exotic matter inside NS is a major difficulty, e.g., 
it is hard to distinguish between a purely hadronic NS and one with a quark core solely from NS observables, such as the star mass, radius and tidal deformability.
However, the presence of a first-order phase transition to quark matter can imprint signatures in NS binary mergers observations, favoring the hypothesis of quark matter in the NS core \cite{Most:2018eaw,Alford:2019oge,Weih:2019xvw}.\\

QCD effective models that incorporate important properties and symmetries of the strong interaction are widely used for describing quark degrees of freedom in NS matter.
The Nambu-Jona--Lasinio (NJL) model is one of such effective models that considers chiral symmetry preserving interactions \cite{Hatsuda:1994pi,Buballa:2003qv}, and that has been used to study the possible existence of quark matter inside NS \cite{Schertler1999,Hanauske:2001nc,Baldo2002,Menezes2003,Pagliara:2007ph,Bonanno:2011ch,Lenzi:2012xz,Masuda:2012ed,Klahn:2013kga,Logoteta:2013ipa}.

The description of NS matter with non-nucleonic degrees of freedom is
achieved by constructing a hybrid EoS. A possible method to obtain the
EoS is to consider a two-model approach: one that describes the hadronic (confined) phase and a
second model describing  the quark (deconfined) phase. 
While different procedures can be used for the matching of the two EoS \cite{Glendenning2012}, 
we consider a Maxwell construction in order to describe a first-order phase transition from
hadron to quark mater. The reliability of such approach
depends on the still unknown surface tension between both phases,
however, it is justified and has been widely considered
\cite{Pagliara:2007ph,Benic:2014iaa,Benic:2014jia,Zacchi:2015oma,Pereira:2016dfg,Wu:2018kww},
if the surface tension, $\sigma$, is of the order of
$\sigma\gtrsim 40$ MeV fm$^{-2}$ \cite{Yasutake2014}. 
The NJL model has been explored in describing the quark phase of a hybrid EoS in several works,
e.g., \cite{Pagliara2007,Bonanno:2011ch,Pereira:2016dfg}. 
The presence of the vector-isoscalar interactions in the NJL model was shown to be very important in stiffening the EoS to sustain $2M_\odot$. 
The 8-quark  interactions were first introduced to stabilize the NJL
vacuum  of the four-quark NJL and six-quark 't Hooft
interactions \cite{Osipov2005,Osipov2006}. The study of the effect of
scalar and vector 8-quark terms in stellar matter was done in \cite{Benic:2014iaa}, where it
was shown that these terms allow for the description of two solar mass
hybrid  stars, with the 8-quark scalar channel giving rise to a
reduction of the onset of quark matter, and the 8-quark vector channel
stiffening the equation of state at high densities.  Within this same
model it was possible to describe twin stars \cite{Benic:2014jia}
considering a very stiff hadronic EoS. In
\cite{Alvarez2016a,Alvarez2016b} further studies were undertaken with
the same model, in particular, a Bayesian analysis which studied the
possibility of discriminating a  hybrid EoS with a strong first order
phase transition over other possible EoS \cite{Alvarez2016a}. It was also studied the
possible existence of a critical end point in the QCD phase diagram,
taking into account the 2$M_\odot$ NS constraint \cite{Alvarez2016b}.

Local and non-local versions of
NJL models including the 4-quark vector interaction typically would
not predict the existence of hybrid stars (or just small quark
branches) \cite{Schertler1999,Menezes2003,Ranea-Sandoval:2015ldr}.
Other approaches for describing quark matter in NS were also explored. 
In \cite{Alford:2013aca,Alford:2015gna}, the high-density EoS region was analyzed by a constant speed of sound parametrization, showing that a 
large speed of sound in the quark phase, $v_s^2\gtrsim 0.5$ for soft hadronic 
EoS and $v_s^2\gtrsim0.4$ for stiff hadronic EoS would be required for $M_{\text{max}}>2M_{\odot}$.
Using the same formalism, \cite{Han:2019bub}  showed that strong
repulsive interactions in quark matter are required to support the NS masses $M\gtrsim 2.0M_{\odot}$. 
In \cite{Annala2020}, imposing constraints from observational and theoretical ab-initio calculations,
the authors proposed that 1.4$M_\odot$ NS are compatible with hadronic stars, by using a speed of sound EoS parametrization. 
Furthermore, the authors inferred that massive
stars $\approx 2M_\odot$ and a  speed of sound 
not far from  the conformal limit will have large quark cores. \\

In our previous works \cite{Ferreira:2020evu,Ferreira:2020kvu}, we have analyzed hybrid stars using the NJL model with different interaction terms at the Lagrangian level which describes the quark phase. 
The usual three-flavor NJL model \cite{Klevansky1992,Hatsuda:1994pi} with additional vector and pseudovector interactions and the vector-isovector and
pseudovector-isovector interactions were explored in
\cite{Ferreira:2020evu}. It was shown that considerable quark core
sizes would require moderate values for the quark vector-isoscalar term and a weak  vector isovector term.
Additional higher-order repulsive interactions,
4-quark and 8-quark  vector-isoscalar interactions,
were included in \cite{Ferreira:2020kvu}.
The 8-quark vector-isoscalar channel was shown to allow for the appearance of a quark core at moderately low NS masses, $\sim 1M_{\odot}$, while providing the required repulsion to keep the star stability up to $\sim2.1M_{\odot}$. 
It was also shown that both the heaviest NS mass and radius are  sensitive to the strength of 8-quark  vector-isoscalar channel \cite{Ferreira:2020kvu}.\\

The present work is an extension of our
previous works: we  explore the effect of all
the 4-quark and 8-quark vector
interaction channels within the three-flavor NJL model in describing
hybrid EoS .
We will  investigate the impact of the interaction channels in the
existence and stability of hybrid star sequences and on the quark core
properties, such as mass and radius. The possibility of having quark
cores in light NS and fulfilling all observational constraints will be
studied.  In particular, we will look at the strangeness content and
analyse whether NS with large quark cores may have a high $s$-quark
fraction. Immediately after the measurement of the mass of the pulsar
PSR J1614-2230, it was suggested that exotic
degrees of freedom such as strangeness might be ruled out with the
detection of a two
solar mass NS \cite{Demorest2010,Vidana2010}. However, since then, several studies have
shown that in fact it is possible that hyperons nucleate inside a star
with at least two solar masses  \cite{Bednarek2011,Weissenborn2011,Weissenborn2011a,Providencia2012,Oertel2014,Fortin2016}, see \cite{Chatterjee2015} for
a review. Also inside hybrid stars, the existence of strange quarks
was not ruled out by the two solar mass constraint, see for instance
\cite{Bonanno:2011ch,Pereira:2016dfg,Dexheimer2020}. Recently,  within
a Bayesian analysis, which incorporates both information from  the
GW170817 binary neutron star merger \cite{TheLIGOScientific:2017qsa,PhysRevLett.121.161101}, the MP J0740+6620
\cite{Cromartie2019} and the pulsar PSR J0030+0451 as detected by
NICER \cite{Riley_2019}, it was shown that maximum mass of an hybrid star
could be above 2$M_\odot$ \cite{LI2021}.
 
This paper is organized as follows: in Section \ref{model_and_formalism} the quark model is presented. 
The results are discussed in Section \ref{results} and our conclusions
are drawn in Section \ref{conclusions}.

\section{Model and Formalism}
\label{model_and_formalism}
For the quark matter, we consider the following multi-quark interaction Lagrangian density

\begin{widetext}
\begin{align*}
\mathcal{L} =  
\bar{\psi} 
(
i\slashed{\partial} - \hat{m} + \hat{\mu} \gamma^0 
) 
\psi 
& + G_S  \sum_{a=0}^8
[ (\bar{\psi} \lambda^a \psi)^2 + 
(\bar{\psi} i \gamma^5 \lambda^a \psi)^2 ]
 - G_D [  
\det( \bar{\psi} (1+\gamma_5) \psi ) + 
\det( \bar{\psi} (1-\gamma_5) \psi )  ]
\\
& - G_\omega [ (\bar{\psi}\gamma^\mu\lambda^0\psi)^2 + (\bar{\psi}\gamma^\mu\gamma_5\lambda^0\psi)^2 ]
 - G_\rho \sum_{a=1}^8 
[ 
(\bar{\psi} \gamma^\mu\lambda^a \psi)^2 +  
(\bar{\psi} \gamma^\mu\gamma_5\lambda^a \psi)^2 
]\\
&
- G_{\omega \omega} 
[ (\bar{\psi}\gamma^\mu\lambda^0\psi)^2 + (\bar{\psi}\gamma^\mu\gamma_5\lambda^0\psi)^2 ]^2
\\
& - G_{\rho \rho} \sum_{a=1}^8 \sum_{b=1}^8 
[ 
(\bar{\psi} \gamma^\mu\lambda^a \psi)^2 +  
(\bar{\psi} \gamma^\mu\gamma_5\lambda^a \psi)^2 
]
[ 
(\bar{\psi} \gamma^\mu\lambda^b \psi)^2 +  
(\bar{\psi} \gamma^\mu\gamma_5\lambda^b \psi)^2 
]
\\
& - G_{\omega \rho} 
[ (\bar{\psi}\gamma^\mu\lambda^0\psi)^2 + (\bar{\psi}\gamma^\mu\gamma_5\lambda^0\psi)^2 ]
\sum_{a=1}^8 
[ 
(\bar{\psi} \gamma^\mu\lambda^a \psi)^2 +  
(\bar{\psi} \gamma^\mu\gamma_5\lambda^a \psi)^2 
].\numberthis
\label{eq:SU3_NJL_lagrangian}
\end{align*} 
\end{widetext}
This is a SU$(3)$ NJL-type model that includes four and six
scalar-pseudoscalar interactions, and four and eight vector interactions.
The quark current masses and chemical potentials are given by $\hat{m}=\diag(m_u, m_d, m_s )$ and $\hat{\mu}=\diag(\mu_u, \mu_d, \mu_s )$, respectively.
The model is regularized by a 3-momentum cutoff $\Lambda$. We note
  that the  Lagrangian density  terms designated
with a subscript $\rho$ involve  the three flavors, $u$, $d$ and $s$ and,
therefore, are hypercharge
operators.\\

The four scalar and pseudoscalar quark interaction  are present in the original formulation of the NJL model and are essential to incorporate spontaneous chiral symmetry breaking in the model. The 't Hooft determinant for three quark flavors corresponds to a six quark interaction which incorporates the explicit $U_A(1)$ symmetry breaking in the model. Incorporating vector interaction in the model has been found to be necessary to model the medium to high density behavior of the EoS and predict $2M_\odot$ NSs. The inclusion of all possible chiral-symmetric set of eight quark vector interactions was performed in \cite{Morais:2017bmt}, in order to study the masses of the lowest spin-0 and spin-1 meson states. 

In the mean-field approximation, the model thermodynamical potential is given by  
\begin{widetext}
\begin{align*}
\Omega - \Omega_0 & = 
2 G_S  
(  
\sigma_u^2 + \sigma_d^2 + \sigma_s^2 
) 
- 4 G_D \sigma_u\sigma_d \sigma_s 
\\
& 
- \frac{2}{3} G_\omega
( 
\rho_u + \rho_d + \rho_s 
)^2  
- G_\rho
( 
( \rho_u - \rho_d )^2 +
\frac{1}{3}
( \rho_u + \rho_d - 2\rho_s )^2
)
\\
& 
-\frac{4}{3 }G_{\omega \omega}
( 
\rho_u + \rho_d + \rho_s 
)^4
- 3 G_{\rho \rho}
( 
( \rho_u - \rho_d )^2 +
\frac{1}{3}( \rho_u + \rho_d - 2\rho_s )^2
)^2
\\ 
&
- 2 G_{\omega \rho}
( 
\rho_u + \rho_d + \rho_s 
)^2  
(
( \rho_u - \rho_d )^2 +
\frac{1}{3}( \rho_u + \rho_d - 2\rho_s )^2
)
\\  
& - 2 N_c \sum_{i=u,d,s} 
\int \frac{ d^3p }{(2\pi)^3}   
[  
E_i + 
T \ln ( 1 + e^{-(E_i+\tilde{\mu}_i)/T} ) + 
T \ln ( 1 + e^{-(E_i-\tilde{\mu}_i)/T}  )
], \numberthis
\label{NJLpot}
\end{align*}
\end{widetext}
where $E_i=\sqrt{p^2+M_i^2}$, $\sigma_i$ and $\rho_i$ are the $i$-flavor quark condensate and quark density.
The constant $\Omega_0$ is fixed to give a vanishing vacuum potential. The values for the quark condensates, $\sigma_i$, and densities, $\rho_i$, are obtained by requiring the potential to be stationary with respect to the effective masses and effective chemical potentials, $\partial \Omega / \partial M_i = 0$ and $\partial \Omega / \partial \tilde{\mu}_i=0$ i.e., imposing thermodynamic consistency \cite{Buballa:2003qv}. In the mean-field approximation, the product between quark bilinear operators in Eq. (\ref{eq:SU3_NJL_lagrangian}) is linearized, providing an effective Lagrangian which is quadratic in the fermion fields. Using such approximation, it is possible to obtain the thermodynamical potential written in Eq. (\ref{NJLpot}). For more details on the linear expansion of the product between $N$ operators, see \cite{CamaraPereira:2020rtu}. 
For $i\ne j\ne k\in \{u,d,s\}$, the effective mass, $M_i$, and effective chemical potentials, $\tilde{\mu}_i$, are given by:
\begin{align*}
M_i = m_i 
& - 4G_S \sigma_i
+ 2G_D \sigma_j \sigma_k ,\numberthis \\ 
\tilde{\mu}_i = 
\mu_i 
& - 
\frac{4}{3} 
G_\omega 
( 
\rho_i+\rho_j+\rho_k
)
\\
& - 
\frac{4}{3} 
G_\rho 
( 
2\rho_i-\rho_j-\rho_k
)
\\
& - 
\frac{16}{9} 
G_{\omega \omega} 
( 
\rho_i+\rho_j+\rho_k
)^3 .
\\
&
-\frac{32}{9} G_{\rho \rho}  
(
2 \rho_i -\rho_j - \rho_k
)^3 
\\
&
-
\frac{32}{3} G_{\rho \rho} 
(
\rho_i - \rho_j
) 
(
\rho_k - \rho_i
) 
(
2 \rho_i - \rho_j - \rho_k
) 
\\
&
-
\frac{8}{9} G_{\omega \rho} 
(
\rho_i + \rho_j + \rho_k
)\\
&\times
(
4 \rho_i^2 + \rho_j^2 + \rho_k^2 
- \rho_i \rho_j   - \rho_i \rho_k - 4 \rho_j \rho_k 
)
.
\numberthis
\end{align*}

By taking the zero temperature limit in Eq. (\ref{NJLpot}), one can determine the pressure and energy density for cold quark matter
$P = -\Omega $ and $\epsilon  = -P +  \sum_i \mu_i \rho_i$, respectively. In the following analysis, we will represent the EoS quantities as a function of baryonic density, $n$, which is given by $n=(\rho_u+\rho_d+\rho_s)/3$.\\

The model interaction couplings include the usual NJL parameters
$\{G_S,G_D,\Lambda \}$ and the following additional extra terms
$\{G_\omega,G_{\omega \omega},G_{\rho}, G_{\rho \rho}, G_{\omega \rho}
\}$. Our goal is to analyze the overall effect of the additional terms
on the NS properties  of hybrid EoS. For that, we fix $G_S$, $G_D$,
and $\Lambda$ to reproduce the meson masses
of the $\pi^{\pm}$, $K^{\pm}$,  $\eta$ and $\eta'$
and the leptonic decay constants of the  $\pi^{\pm}$ and $K^{\pm}$,  $f_{\pi^{\pm}}$ and $f_{K^{\pm}}$, while leaving the extra coupling  as free parameters.
We show the parameter set used in Table \ref{tab:2}, and the model predictions, within the present parametrization, for some meson masses and leptonic decay constants in Table \ref{tab:3}.

\begin{table}[!htb]
\begin{tabular}{cccccccc}
    \hline
    \hline
$\Lambda$  & $m_{u,d}$ & $m_s$  & $G_S\Lambda^2 $ & $G_D\Lambda^5 $ & $M_{u,d}$ & $M_s$ \\
\text{[MeV]}      &  [MeV]    & [MeV]   &               &                  & [MeV] &  [MeV]\\
\hline
   \hline
623.58 & 5.70   & 136.60 & 1.67 &  13.67 & 332.2   & 510.7 \\
   \hline
\end{tabular}
\caption{Parameters of the NJL  model used in the present work: $\Lambda$ is the model cutoff, $m_{u,d}$ and $m_{s}$ are the quark current masses, $G_S$ and $G_D$ are coupling constants. $M_{u,d}$ and $M_{s}$ are the resulting constituent quark masses in the vacuum. This parameter set yields, in the vacuum, a light quark condensate of $\ev{\bar{q}_l q_l}^{1/3}=-243.9~\mathrm{MeV}$  and strange quark condensate of $\ev{\bar{q}_s q_s}^{1/3}=-262.9~\mathrm{MeV}$.}
\label{tab:2}
\end{table}

\begin{table}[!htb]
    \begin{tabular}{ccc}
    \hline
    \hline
     & NJL SU(3) {   }& Experimental \cite{Agashe:2014kda} \\
        \hline
          \hline
    $m_{\pi^{\pm}}$ [MeV]     & 139.6  & 139.6 \\
    $f_{\pi^{\pm}}$ [MeV]     & 92.0   & 92.2 \\
    $m_{K^{\pm}}$ [MeV]       & 493.7  & 493.7 \\
    $f_{K^{\pm}}$ [MeV]       & 96.4   & 110.4 \\
    $m_{\eta}$ [MeV]          & 515.6  & 547.9 \\
    $m_{\eta'}$ [MeV]         & 957.8  & 957.8 \\
    \hline
    \end{tabular}  
    \caption{The  masses and decay constants of several mesons within the model and the respective experimental values. }
  \label{tab:3}
\end{table}

As in our previous works \cite{Pereira:2016dfg}, each model  parametrization is identified by dimensionless ratios instead of the actual couplings values. The quark models are characterized by the following five coupling ratios: $\chi_{\omega} = G_\omega / G_S$, $\chi_{\omega \omega} = G_{\omega \omega} / G_S^4$, $\chi_{\rho} = G_\rho / G_S$, $\chi_{\rho \rho} = G_{\rho \rho} / G_S^4$, and $\chi_{\omega \rho} = G_{\omega \rho} / G_S^4$. In the following, we identify each model by the set 
of values $\{\chi_{\omega},\chi_{\omega \omega}, \chi_{\rho}, \chi_{\rho \rho}, \chi_{\omega \rho} \}$. The model pressure, $P$, and energy density, $\epsilon$, are defined up to an extra constant term $B$, i.e.,  $P \to P + B$ and $\epsilon \to \epsilon-B$. 
The effect of the bag parameter $B$ was already widely studied \cite{Hanauske:2001nc,Klahn:2006iw,Pagliara:2007ph,Bonanno:2011ch,Lenzi:2012xz,Masuda:2012ed,Klahn:2013kga,Logoteta:2013ipa,Menezes:2014aka,Klahn:2015mfa,Pereira:2016dfg,Ferreira:2020evu}. It was found that the baryonic density at which 
the onset of quark matter occurs decreases with increasing $B$.
We have verified that this behavior remains the same in the presence of any 
of the considered interactions and regardless of their coupling values.
Therefore, for the sake of simplicity, we fix, hereafter, the bag parameter 
to $B=15$ MeV/fm$^{3}$ \footnote{There was a misprint in the $B$ value reported in our previous work \cite{Ferreira:2020kvu},  
being $B=15$ MeV/fm$^{3}$ the value used instead of the reported value, $B=10$ MeV/fm$^{3}$}.\\

Each hybrid EoS consists of a hadronic phase connected to a quark phase through a first-order phase transition. This two-model approach has been widely used \cite{Pagliara:2007ph,Benic:2014iaa,Benic:2014jia,Zacchi:2015oma,Pereira:2016dfg,Wu:2018kww}. The first-order phase transition from hadronic to quark matter is achieved imposing the Maxwell construction, in which both phases are in chemical, thermal and mechanical equilibrium: $\mu_B^H = \mu_B^Q$, $P^H = P^Q$, and $T^H = T^Q$,
where $\mu_B$ is the baryon chemical potential, $P$ the pressure and the labels $H$ and $Q$ represent the hadronic and quark phases, respectively.
The DDME2 model is used for the hadronic part \cite{ddme2}. This is a
relativistic mean-field model with density dependent couplings  that
describes two solar mass stars and  satisfies a well established set
of nuclear matter and finite nuclei constraints
\cite{Dutra2014,Fortin2016}, including the constraints set by the
ab-initio calculations for neutron matter using a chiral effective
field theoretical approach \cite{Hebeler2013}.  
It is important to realize that the hadronic EoS also has an impact in the hybrid star EoS: having a softer EoS, for example, would shift
the deconfinement transition to larger densities, giving rise to smaller quark branches, and to the possibility of not fulfilling the 2$M_\odot$ star constraint. The opposite would be valid for a
harder hadronic EoS. In \cite{Ferreira:2020evu} we have shown that, in order to 
get 2$M_\odot$  hybrid stars, the hadronic EoS has to be in average stiffer
than necessary for a 2$M_\odot$  hadronic star.  Having this in mind,
we chose for the hadronic phase an EoS that would allow 2 $M_\odot$
hybrid stars and still satisfy well established nuclear properties.

\section{Results}
\label{results}
The aim of the present work is to analyze the effect of quark matter on the hybrid stars properties. The quark model we are considering  has five free parameters: $\chi_{\omega} = G_\omega / G_S$, $\chi_{\omega \omega} = G_{\omega \omega} / G_S^4$, $\chi_{\rho} = G_\rho / G_S$, $\chi_{\rho \rho} = G_{\rho \rho} / G_S^4$, $\chi_{\omega \rho} = G_{\omega \rho} / G_S^4$. 
In our previous works \cite{Ferreira:2020evu,Ferreira:2020kvu}, we have already analyzed some interactions of the present quark model (Eq. \ref{eq:SU3_NJL_lagrangian}), which we briefly summarize. The couplings  $\{\chi_{\omega}, \chi_{\rho} \}$ were analyzed in \cite{Ferreira:2020evu}, where a set of hadronic EoS was used. In there, the $\chi_\omega$ coupling was seen to stiffen the quark EoS, with the onset
of quarks occurring at larger densities, giving rise to more massive
hybrid stars with  smaller quark cores. Similar results were also
discussed in \cite{Logoteta:2013ipa,Pereira:2016dfg,Ferreira:2020kvu}.  
On the other hand, $\chi_\rho$ softens the quark EoS at higher densities, as the onset of the strange quarks is pushed to lower densities, but stiffens the
quark EoS at lower densities while the $s$-quark does not set in. As a result, the
onset of quark matter occurs at higher densities for a non-zero $\chi_\rho$, and sizeable quark cores are only obtained if $\chi_\rho<0.4$ \cite{Pereira:2016dfg,Ferreira:2020evu}. The effect of the
$\{\chi_{\omega},\chi_{\omega \omega} \}$ interactions was the purpose
of our work \cite{Ferreira:2020kvu}. Besides the effect of
$\chi_{\omega}$ we just referred, it was also shown that 
$\chi_{\omega \omega}$ induces a non-linear density
dependence of the speed of sound, which is crucial to generate
large quark cores. \\

Despite controlling the onset density of strange quarks, the impact of both $\chi_{\rho \rho}$ and $\chi_{\omega \rho}$ couplings on the EoS quark matter in $\beta$-equilibrium is weak when compared with the other couplings. 
Almost all NS properties show small changes with both couplings and,
for the sake of simplicity, we fix them to  $\chi_{\rho
  \rho}=\chi_{\omega\rho}=0$ hereafter.
While a detailed study of both couplings is left for future work, we
give later a brief overview of their impact. 
The effect of $\chi_{\omega}$ was confirmed, i.e., the onset of quark matter happens at larger densities
with  increasing coupling, giving rise to smaller quark cores. 
In a first approach, we consider $\chi_{\omega}=0$, and analyze the impact of  $\{ \chi_{\omega \omega},  \chi_{\rho}\}$ on the properties of hybrid stars.
In a second step, we consider the interplay among the three
couplings,   $\{ \chi_{\omega \omega}, \,  \chi_{\rho}, \, \chi_{\omega}\}$.
In fact, as expected, a non-zero $\chi_{\omega}$ 
changes the results as just referred, e.g. a finite
$\chi_{\omega}$ gives more massive stars with
smaller quark cores.\\

\subsection{Quark matter EOS}

\begin{figure}[!htb] 
	\centering
    \includegraphics[width=1.0\columnwidth]{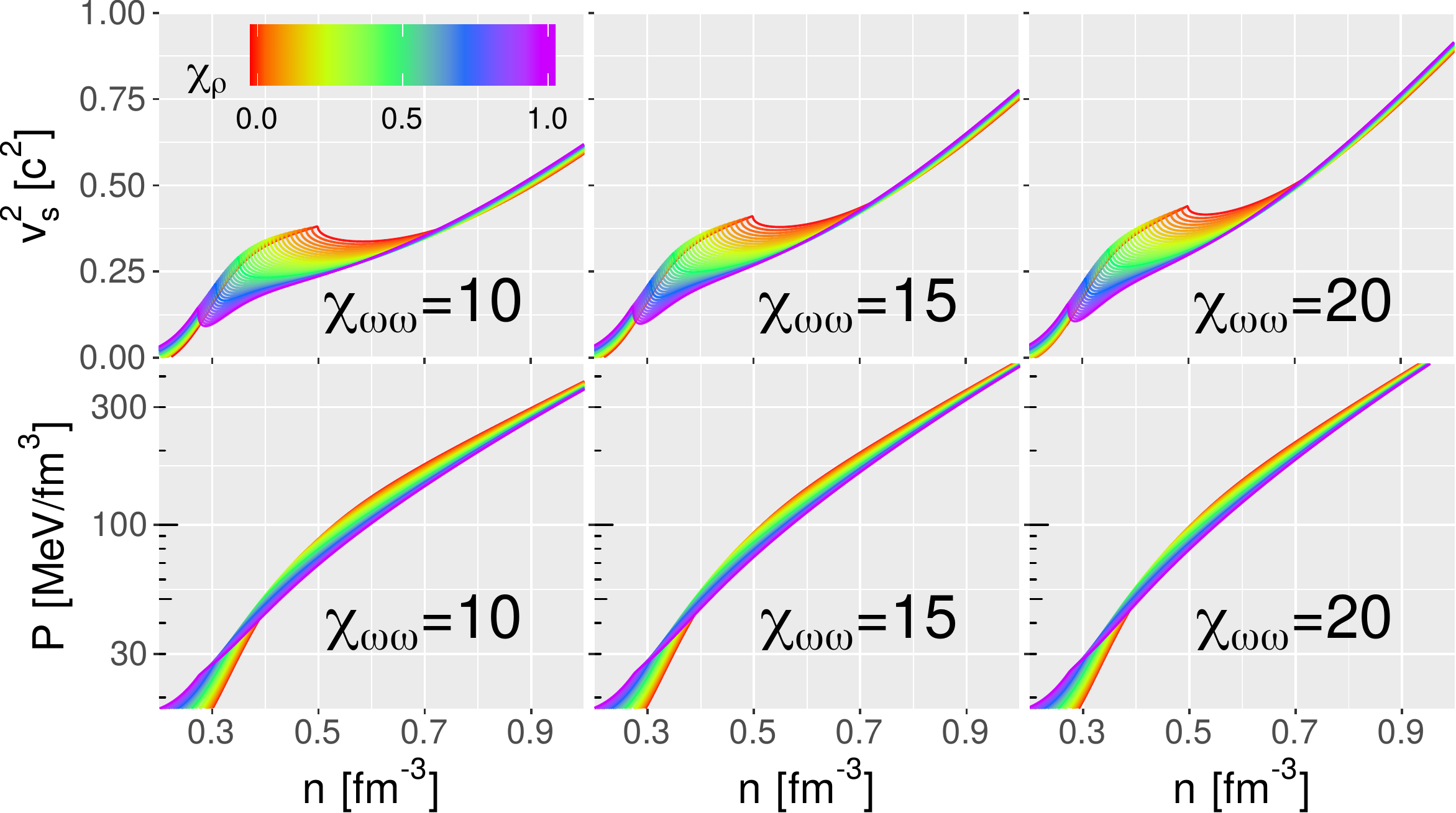}
	\caption{Quark matter squared speed of sound (top) and pressure (bottom) as function of baryon density for different values of 	$\chi_{\rho}$ (color scale) with $\chi_{\omega\omega}=10$ (left),
	$\chi_{\omega\omega}=15$ (center), and $\chi_{\omega\omega}=20$ (right).}
\label{fig:fig1}
\end{figure}

We first discuss the joint effect of the couplings $\{ \chi_{\omega\omega},  \chi_{\rho} \}$ on the properties of the EoS, and set $\chi_{\omega}=0$.
Figure \ref{fig:fig1} shows the squared speed of sound (top panels) and the pressure (bottom panels) of the quark EoS as a function of baryonic
density for different  values of $\chi_{\rho}$ (color scale) and
$\chi_{\omega \omega}$ (different panels).
As $\chi_{\omega \omega}$ increases, the pressure, $P(n)$, increases at high densities: the EoS  clearly  shifts to larger values of $P$. 
The increase of  the $\chi_{\rho}$ coupling shifts the onset of
strange quarks to lower densities with direct consequences on the
stiffness of the EoS: a) at low densities, below the $s$-quark onset, the
larger the value of  $\chi_{\rho}$ the stiffer the EoS; b) the EoS smooths as soon as the $s$-quark sets in, and this happens
first to the large values  of  $\chi_{\rho}$ (purple color),
and above this onset density the EoS with larger values of
$\chi_{\rho}$ have the smaller pressures. 
The $v_s^2(n)$  reflects this behavior:
the sudden decrease between 0.3 and 0.5 fm$^{-3}$ signals the
appearance of strange quarks which
moves towards lower densities with increasing  $\chi_{\rho}$. A smoother
crossover to strange matter is obtained with smaller values of $\chi_{\rho}$
because the fraction of $s$-quarks increases more slowly, see \cite{Pereira:2016dfg}.

For the highest densities, $v_s^2(n)$ is again  larger for the
larger $\chi_{\rho}$ couplings, revealing the stiffening effect of
$\chi_{\rho}$ once the fraction of $s$ quarks becomes equilibrated.

\subsection{NS properties}

\begin{figure}[!htb] 
	\centering
	\includegraphics[width=1.0\columnwidth]{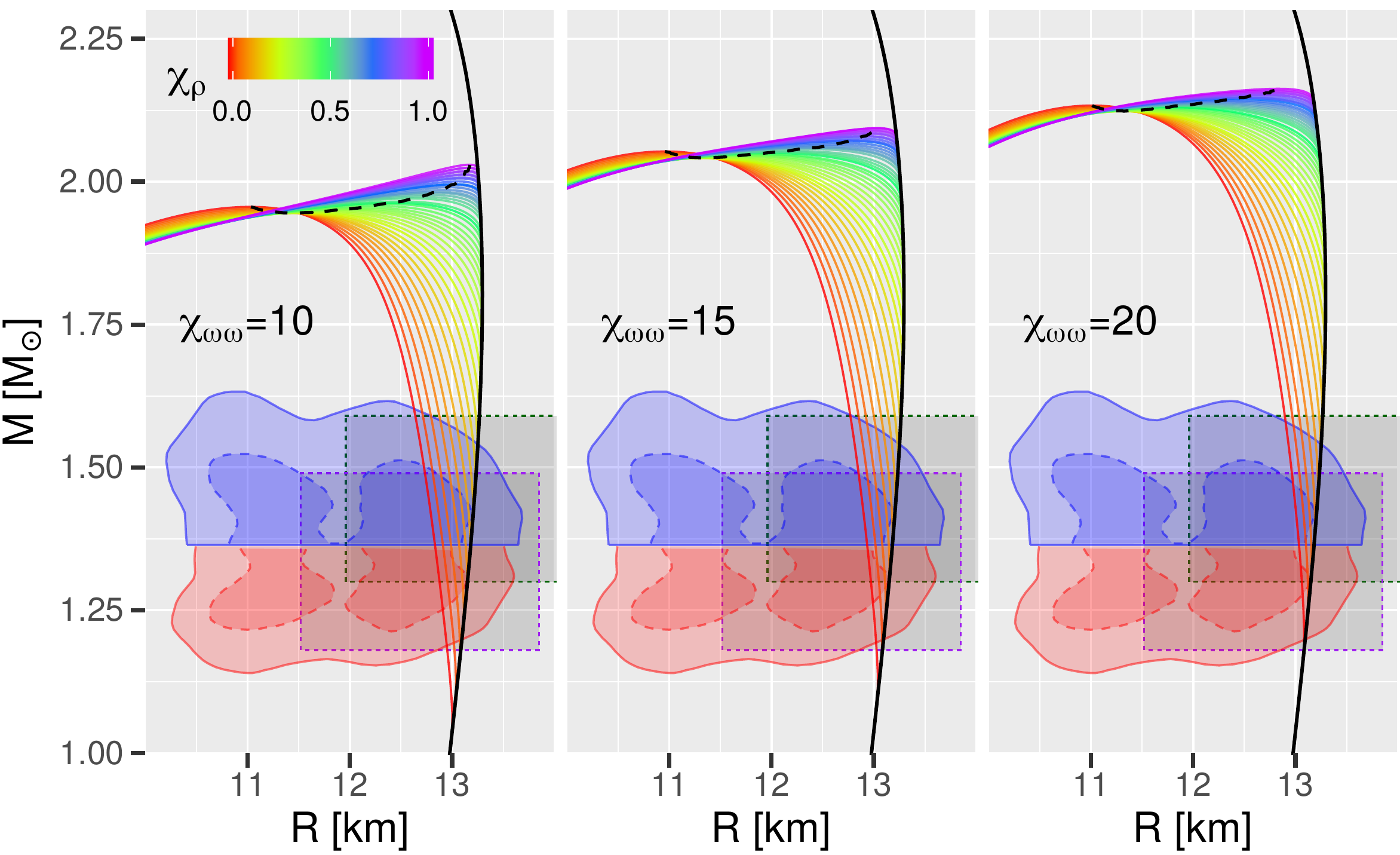}
	\caption{$M(R)$ diagrams for $\chi_{\omega\omega}=10$ (left), $\chi_{\omega\omega}=15$ (center), and 
	$\chi_{\omega\omega}=20$ (right). The color scale indicates the $\chi_{\rho}$ value and the solid black line represents the purely hadronic sequence. The black dashed line indicates the maximum mass reached by each EoS. The top blue and bottom red regions indicate, respectively, the 90\% (solid) and 50\% (dashed) credible intervals of the LIGO/Virgo analysis for each binary component from the GW170817 event \cite{Abbott:2018exr} (using a set of parametrized EoS that assumes $M_{\text{max}}\geq1.97 M_{\odot}$). The rectangular regions enclosed by dotted lines indicate the constraints from the millisecond pulsar PSR J0030+0451 NICER x-ray data  \cite{Riley_2019,Miller:2019cac}.}
	\label{fig:TOV}
\end{figure}

In order to study NS properties, we have integrated the Tolmann-Oppenheimer-Volkoff (TOV)
equations \cite{TOV1,TOV2}, together with the differential equations that determine
the tidal deformability \cite{Hinderer:2009ca}. 
The sequence of stars for each hybrid EoS, parametrized by $\{
\chi_{\omega \omega},\chi_{\rho} \}$ and $\chi_{\omega} =0$, is presented in
Fig. \ref{fig:TOV}.
Three scenarios are identified: $\chi_{\omega
  \omega}=10$ (left), $\chi_{\omega \omega}=15$ (center), and
$\chi_{\omega \omega}=20$ (right).
As shown in
\cite{Ferreira:2020kvu}, a larger  $\chi_{\omega \omega}$ allows for larger
quark branches, capable of reproducing more massive NS  with
quarks already present inside light NS. If  $\chi_{\omega\omega}=10$
and $\chi_{\omega}=0$, 2$M_\odot$ stars are not attained except
for large values of $\chi_{\rho}$ where the quark branch is very small. 
However, for both   $\chi_{\omega\omega}=15$ and 20, it is
possible to describe 2 $M_\odot$ stars with large quark branches.
Figure \ref{fig:TOV} also shows the
impact of $\chi_{\rho}$ on the onset of quarks: independently of
$\chi_{\omega\omega}$, the larger $\chi_{\rho}$ the larger the quark
onset star mass, as expected from the discussion of the pressure
behavior.
The second effect of a finite $\chi_{\rho}$  is to reduce
 considerably the quark branch. This is mainly due to the increase in
 the amount of $s$-quarks that smooths the EoS and, therefore, reduces
 the ability to counterbalance gravity. As a consequence,  smaller central densities occur for larger values of  $\chi_{\rho}$, as also discussed  in \cite{Pereira:2016dfg}.
Thus, the radius of the maximum mass, $R_{\text{max}}$, is
an increasing function of  $\chi_{\rho}$. These results are
in accordance with  \cite{Ferreira:2020evu}, where smaller quark cores were generated by larger $\chi_{\rho}$ values. 
It is also interesting to realize that the $\chi_{\rho}$ interaction
is responsible for the crossing of all NS sequences in a small
region of $(M,R)$ diagram. Such crossing is also observed in the
pressure and speed of sound of the model for increasing values of
$\chi_{\rho}$, see Fig. \ref{fig:fig1}. This effect is related to the
decreasing of the  onset density of strange quarks with  increasing $\chi_{\rho}$: increasing $\chi_{\rho}$ stiffens the low density region of the EoS while softening its high density region. This opposite behaviour in the low and high density regions of the EoS, for increasing $\chi_{\rho}$, generates a crossing point in between these densities regimes. Such behaviours is then translated to the mass-radius relations, giving rise to the crossing behaviour observed in Fig. \ref{fig:TOV}.

The set of hybrid EoS fulfill both the
constraints derived from the NICER x-ray data for the millisecond
pulsar PSR J0030+0451 \cite{Riley_2019,Miller:2019cac}   (shaded
rectangular regions)  and the confidence intervals of the posterior probability from 
the LIGO/Virgo analysis \cite{Abbott:2018exr} for the tidal
deformability.
Recent results of NICER \cite{Riley2021,Miller2021}, taking into
  account NICER  and other observations, predict for the pulsar MP
  J0740+6620  the mass and radius
2.08$\pm$0.07 $M_\odot$ and  12.35$\pm$0.75
km \cite{Riley2021}, and 
$2.072_{-0.066}^{+0.067} M_{\odot}$ and   $12.39_{-0.98}^{+1.30}$ km \cite{Miller2021}, at one standard deviation. Besides, the authors of \cite{Riley2021} also predict the  radius
12.45$\pm$0.65 km for a 1.4 $M_\odot$ star, at one standard deviation. Our
results are compatible with these new data.

\begin{figure}[!htb]
	\centering
	\includegraphics[width=1.0\columnwidth]{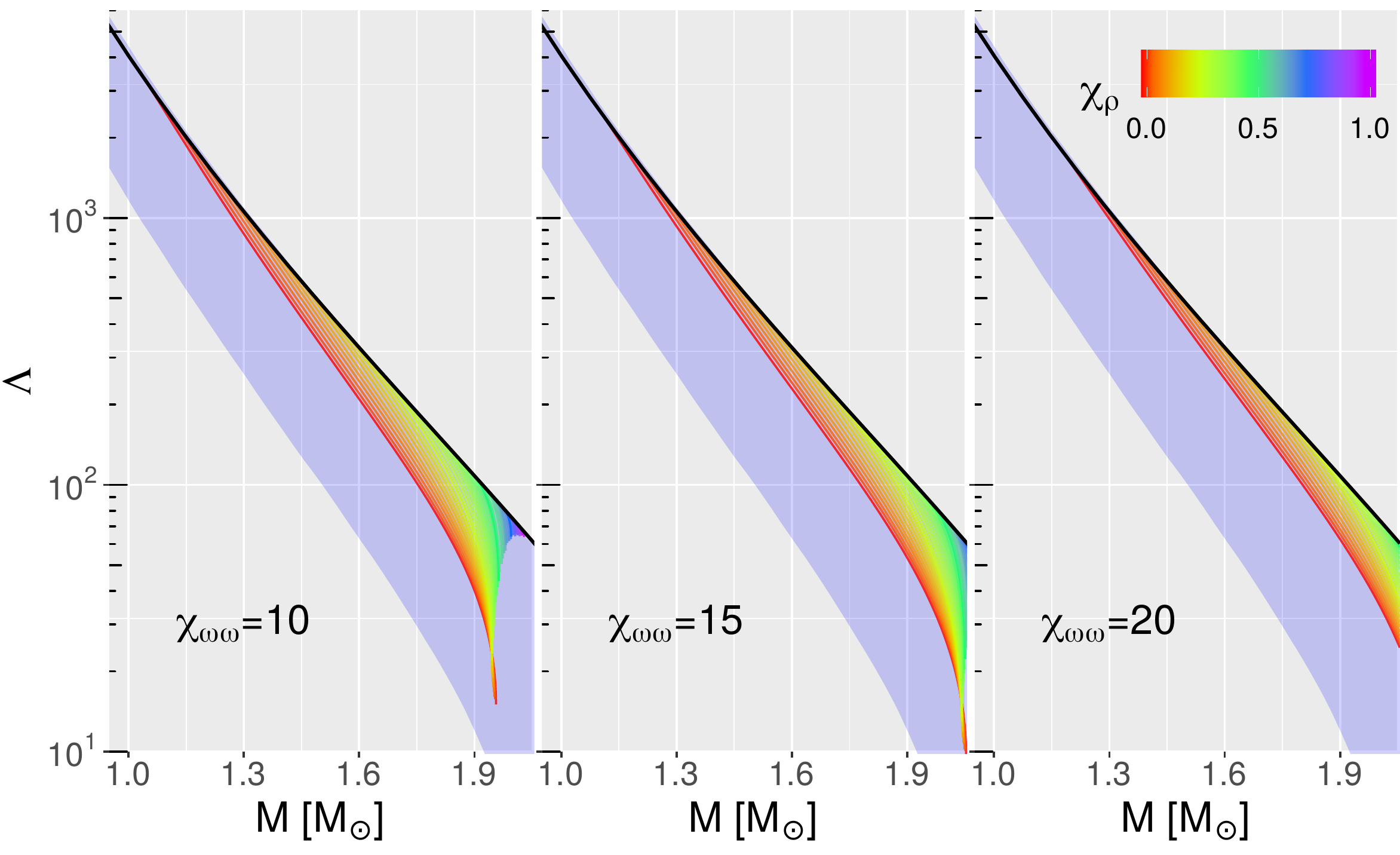}
	\caption{$\Lambda(M)$ diagrams for for $\chi_{\omega\omega}=10$ (left), $\chi_{\omega\omega}=15$ (center), and 
	$\chi_{\omega\omega}=20$ (right). 
	The color scale indicates the $\chi_{\rho}$ value and the
        solid black line represents the purely hadronic sequence.
        The background band is the 90\% posterior credible level obtained in \cite{Abbott:2018exr}, when it is imposed that
          the EoS should describe 1.97$M_\odot$ NS. }
\label{fig:lambda}
\end{figure}

The emitted gravitational-wave signals from
binary NS systems carries information on the  tidal
deformability of a NS, which is an EoS dependent quantity.
For a NS with mass $M$, the dimensionless tidal deformability is given by $\Lambda=2k_2R^5/(3M^5)$,
where $k_2$ is the Love number \cite{Hinderer:2009ca}. 
The $\Lambda(R)$ diagrams are shown in Fig. \ref{fig:lambda}, for the
same couplings defined 
in Fig. \ref{fig:TOV}.  
The background band is the 90\% posterior credible level
obtained in \cite{Abbott:2018exr} when it is imposed that
the EoS should describe 1.97$M_\odot$ stars. 
Only values $\Lambda(R)$ corresponding to hybrid NS with $M_{\text{max}}\geq 1.97M_\odot$ are shown. The $\Lambda(R)$ determined
from the hadron EoS DDME2  lies just at the upper limit of
the gray band. All hybrid stars predict values  of
$\Lambda(R)$ within the band and, as expected,  the following trends are
observed: the larger   $\chi_{\omega\omega}$  and
$\chi_{\rho}$  the larger is  $\Lambda(R)$, and,
therefore, more compressed in the direction of the upper
bound of the LIGO/Virgo 90\% posterior credible level.

\begin{figure}[!htb]
	\centering
	\includegraphics[width=1\columnwidth]{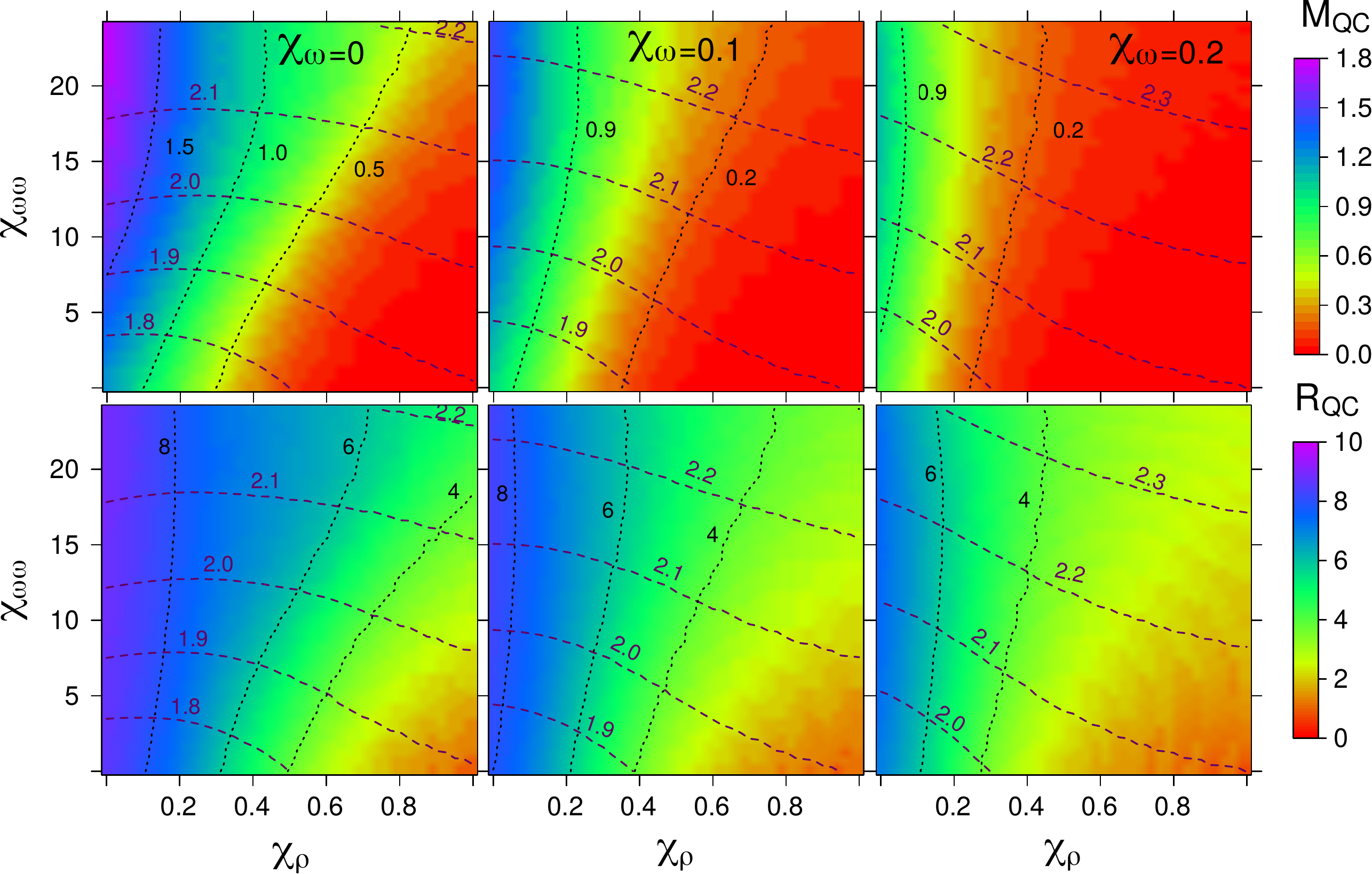}
	\caption{The quark core mass $M_{\text{QC}}$ [$M_{\odot}$] (top)  and  radius $R_{\text{QC}}$ [km] (bottom) as a function of $\chi_{\omega\omega}$	and $\chi_{\rho}$. The brown dashed lines and black dotted lines represent, respectively,  the value of $M_{\text{max}}$ [$M_{\odot}$] and specific $\{M_{\text{QC}},R_{\text{QC}}\}$ contour lines. }
\label{fig:quark_core}
\end{figure}

Let us know investigate how the quark core mass, $M_{\text{QC}}$, and radius, $R_{\text{QC}}$, depend on the couplings $\{ \chi_{\omega \omega},\,\chi_{\rho},\, \chi_{\omega} \}$. 
Figure \ref{fig:quark_core} shows the dependence of $M_{\text{QC}}$ (top) and $R_{\text{QC}}$ (bottom) on the  interactions $\{ \chi_{\omega
\omega},\chi_{\rho} \}$, for $\chi_{\omega}=0, \, 0.1,\, 0.2$,
from  the left to the right, respectively.  The couplings have a
competitive effect: while $ \chi_{\omega \omega}$ increases the
quark core masses up to $1.53M_{\odot}$, a finite value
$\chi_{\rho}$ or/and $ \chi_{\omega} $  has the decreasing
effect. The main conclusions drawn are:  i) if
$\chi_{\omega}=\chi_{\rho}=0$,  2$M_\odot$ NS are only attained
with a value of $ \chi_{\omega \omega}\gtrsim 12$ and very large
quark cores are possible.  As we will see next,
these configurations  correspond to speed of sound in the center of the star close to 1; ii) a finite
$\chi_{\omega}$ or finite $\chi_{\rho}$ allows   2$M_\odot$ NS with
intermediate values of $\chi_{\omega \omega}$, however, for the quark cores  to have at least one third of the total mass of the star $\chi_{\rho}< 0.1 $ if
$\chi_{\omega}=0.2$, and  $\chi_{\rho}\lesssim 0.3 $ if $\chi_{\omega}=0.1$. Similar conclusions are drawn concerning quark core radii above 6  km,
i.e. above half the star radius. This discussion is in agreement
with results in \cite{Ferreira:2020evu} and
\cite{Ferreira:2020kvu}. To summarize, large quark cores require
small values of  $\chi_{\rho}$ and $\chi_{\omega}$ and a large
$\chi_{\omega \omega}$ coupling. We may ask whether the $\chi_{\rho}$
and  the $\chi_{\omega}$  interactions play the same role, or if they
affect the star properties in a different way. As we will see next the
coupling $\chi_{\rho}$ directly affects the amount of strangeness
inside the star.\\

\begin{figure}[!htb]
  \centering
	\includegraphics[width=1\columnwidth]{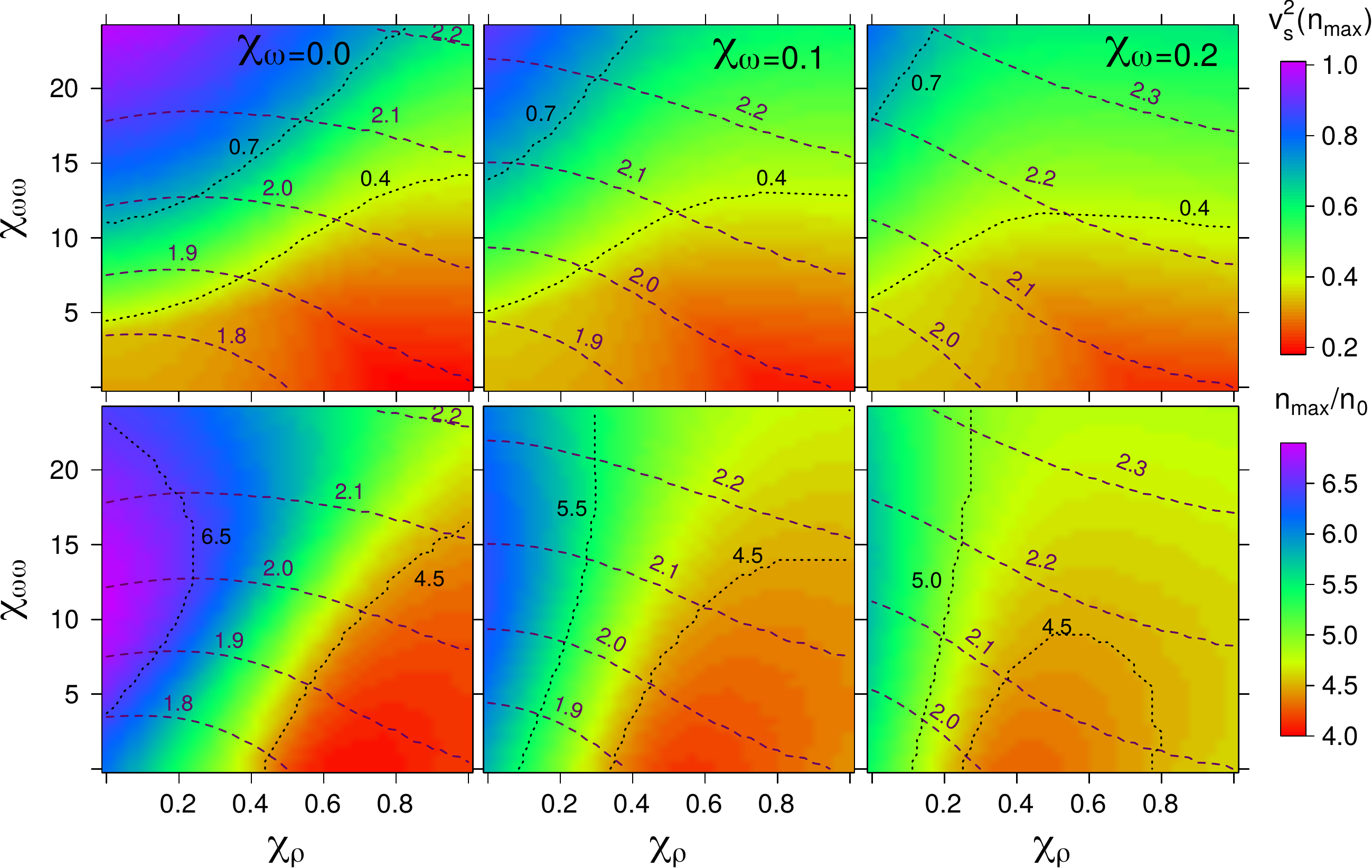}
	\caption{The squared speed of sound [$c^2$] (top)
          and the central densities [$n_0$] of $M_{\text{max}}$ (bottom) as a function of $\chi_{\omega\omega}$	and $\chi_{\rho}$ for $\chi_{\omega}=0$ (left) $0.1$ (center), and $0.2$ (right). $n_0=0.155$ fm$^{-3}$ is the nuclear saturation density.
The brown dashed lines and black dotted lines represent, respectively,  the value of $M_{\text{max}}$ [$M_{\odot}$] and specific $\{v_s^2(n_{\text{max}}),n_{\text{max}}/n_0\}$ contour lines.}
\label{fig:vs_nmax}
\end{figure}

In Fig. \ref{fig:vs_nmax}, we show the $v_s^2(n_{\text{max}})$ (top
panels) and $n_{\text{max}}$ (bottom panels), where $n_{\text{max}}$ is the central density of $M_{\text{max}}$,
as a function of $\chi_{\omega\omega}$	and $\chi_{\rho}$ for three
values of $\chi_{\omega}$: $0$ (left), $0.1$ (center), and $0.2$
(left).
There are  competitive effects from the three couplings on
the $v_s^2(n_{\text{max}})$: 
while $v_s^2$ increases quite rapidly with $\chi_{\omega \omega}$, the
$\chi_{\rho}$ channel smooths this trend. This  also reflects the
fact that a finite value of  $\chi_{\rho}$ reduces
the central density attained. Notice however, that the largest
densities are not attained for the largest  values of
the coupling $\chi_{\omega \omega}$:  for $\chi_{\rho}=0$ the largest central densities
occur with $\chi_{\omega \omega}\sim 12$ and reduce quite fast when
$\chi_{\rho}$ increases; for the largest values of $\chi_{\omega
  \omega}$ considered the  reduction of the central density with
$\chi_{\rho}$ is slower. 
As we will see next, this behavior has an
effect on the strangeness content of the star.
Comparing Figs.  \ref{fig:vs_nmax} and \ref{fig:quark_core}, we
also conclude that the largest quark cores occur for  $\chi_{\omega
  \omega}>17$ when the central densities are smaller than the maximum
value attained for $\chi_{\omega \omega}=12$.
The $\chi_{\omega}$ reduces the range of $v_s^2$ values because it
generates smaller quark cores, see Fig. \ref{fig:quark_core}, and the central densities are not
large enough for  the term $\chi_{\omega\omega}$  to have a strong effect.\\

\begin{figure}[!htb]
	\centering
	\includegraphics[width=1\columnwidth]{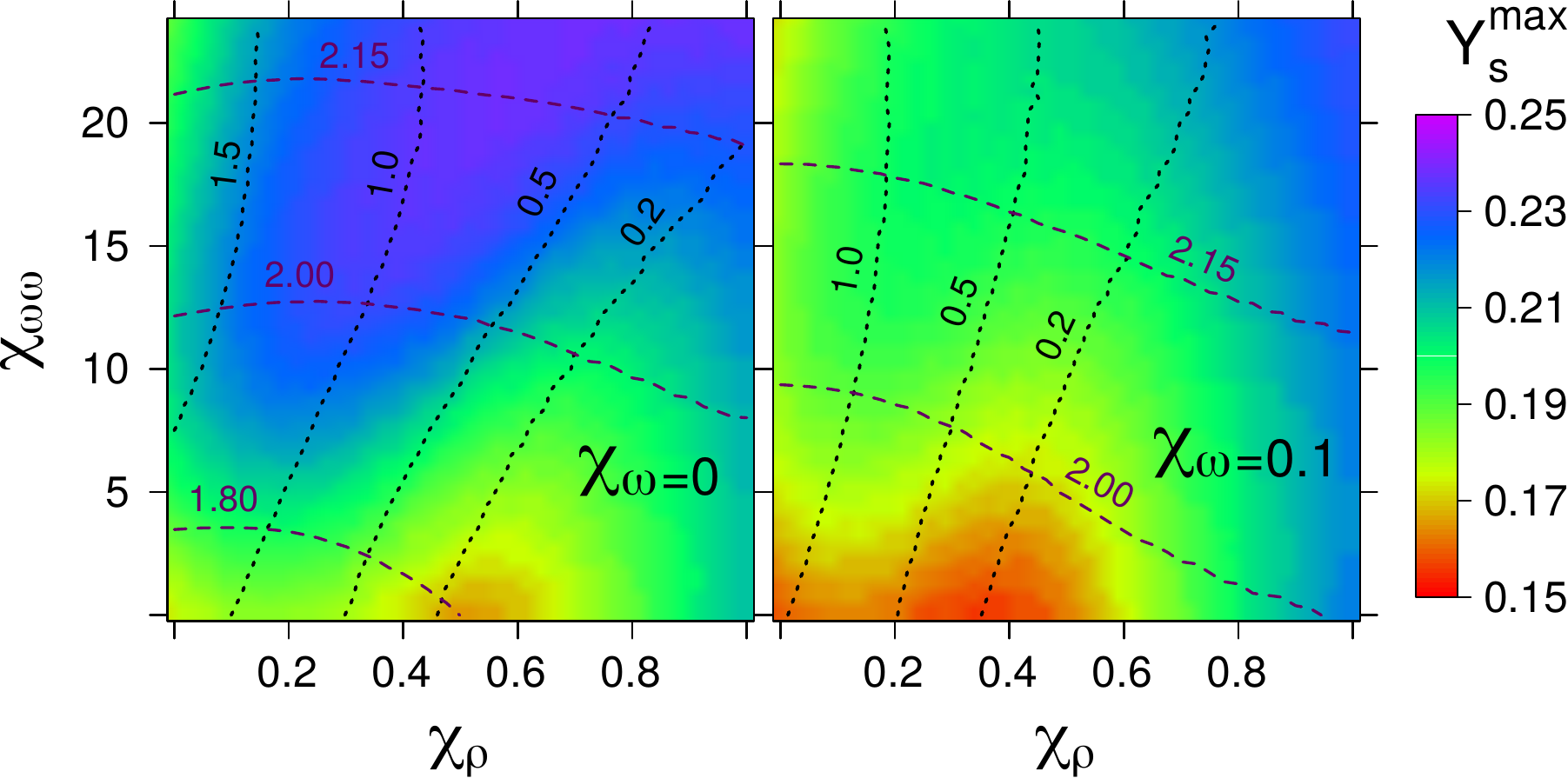}
	\caption{The fraction of $s$-quarks,
          $Y_{s}^{\text{max}}$, at the the central density of $M_{\text{max}}$, $n_{\text{max}}$, as a function of $\chi_{\omega\omega}$	and $\chi_{\rho}$ for $\chi_{\omega}=0$ (left) and $0.1$ (right). 
The brown dashed lines and black dotted lines represent, respectively,  the value of $M_{\text{max}}$ [$M_{\odot}$] and specific $M_{\text{QC}}$ [$M_{\odot}$] contour lines.  }
\label{fig:Y_ds}
\end{figure}

Figure \ref{fig:Y_ds} represents the fraction of $s$-quarks at the central density of $M_{\text{max}}$, 
$Y_{s}^{\text{max}}=\rho^{\text{max}}_{s}/3n_{\text{max}}$.
The $s$-quark fraction shows a non-linear dependence on the couplings
$\chi_{\omega\omega}$ and $\chi_{\rho}$.
We have seen that the larger the value of $\chi_{\rho}$, the
earlier the onset of $s$-quarks. On the other hand, a too large
$\chi_{\rho}$ does not allow a large quark branch and, therefore,
large central densities. Large quark branches are possible including
a strong $\chi_{\omega\omega}$ coupling. The amount of strangeness in the center of the star results from the interplay of these two effects. As a consequence, the largest central fractions of $s$-quarks happen for large $\chi_{\omega\omega}$ and intermediate $\chi_{\rho}$ values, 
while the smallest fractions occur for $\chi_{\omega\omega}=0$: at
finite $\chi_{\rho}$ the $s$-quark sets in earlier and with a finite
$\chi_{\omega\omega}$ larger central densities are attained, and,
therefore, also larger $s$-quark fractions.
The  $\chi_{\omega}$ term reduces the maximum central
density and, therefore,  also the central $s$-quark fraction.  As a
consequence  the effect of  the $\chi_{\omega\omega}$ term is not so
strong.  A large amount of $s$-quarks in the star reduces the quark
core mass, but it is still possible to have hybrid stars with a quark
core with mass of the order of 1/3 to 1/2 of the total star
mass. Moreover, these stars with larger amounts of strangeness 
have squared speed of sound below 1.  Under the conditions studied in the present work the largest fraction of $s$-quark obtained was 25\%, for
$\chi_\rho\approx 0.4$ and $\chi_{\omega\omega}=25$. A larger amount
is prevented because a too large  $\chi_{\omega\omega}$ gives rise
to supra-luminous sound speeds.  The presence of considerably large
fractions of strangeness is essential to allow for phases as the color
flavor locked (CFL)  superconducting phase.

We include in the Appendix two tables with some properties of the
stars obtained with  $\chi_{\omega}=0$ and $\chi_{\omega}=0.1$ and several values of  
$\chi_{\omega\omega}$ and $\chi_{\rho}$, respectively, Tables  \ref{comgea0}  and \ref{comgea1}.
These tables include the quark core ($M_{\text{QC}}$), quark radii ($R_{\text{QC}}$), mass
of the heaviest star ($M_{\text{max}}$), radii of the
heaviest star ($R_{\text{max}}$), onset density of quarks
($n_q$), lightest NS mass with quark content ($M_q$),
central density of the heaviest star ($n_{\text{max}}$),
strange quark fraction at $n_{\text{max}}$
($Y_s^{\text{max}}$), squared speed of sound at
$n_{\text{max}}$ ($v_s^2(n_{\text{max}})$), radii
($R_{1.4M_{\odot}}$) and tidal deformability
($\Lambda_{1.4M_{\odot}}$) of a $1.4M_{\odot}$ NS. Next, we summarize  some of the main conclusions: (i)
quark sets in low mass stars only if $\chi_\omega=0$ and
$\chi_\rho\lesssim 0.2$. Stars with a mass $\approx
1M_\odot$ may have core  quarks; (ii) larger central strangeness
fractions are obtained with larger values of  $\chi_\rho$;
(iii)  the onset density of quarks increases from $\approx
2n_0$ for  $\chi_\rho=0$ to  $\approx
3n_0$ for  $\chi_\rho=0.4$; (iv)  central densities of
maximum mass configuration are not maximum for the largest
values of   $\chi_{\omega\omega}$, but for a given
$\chi_{\rho}$ the maximum mass and maximum $s$-quark central
fraction occurs for the largest $\chi_{\omega\omega}$ values;
(v) the central speed of sound is maximum for the largest
$\chi_{\omega\omega}$ values but decreases when
$\chi_{\rho}$ increases; (vi) even with  $\chi_{\rho}=0.4$ it
is still possible to get a $1 M_\odot$ quark core with a
central density above 6$n_0$; (vii) to attain a maximum mass
configuration $\approx 1.97 M_\odot$
$\chi_{\omega\omega}\gtrsim 10$ .

A finite $\chi_{\rho \rho}$ and/or $\chi_{\omega\rho}$ coupling
would not considerably change the above results. A finite
$\chi_{\rho \rho}$, within the range considered for $\chi_{\omega
\omega}$, i.e., between 0 and 25, would increase the
$Y_{s}^{\text{max}}$ up to 1\%, while both $M_{\text{max}}$ and
$v_s^2(n_\text{max})$ are almost insensitive. A finite value
$\chi_{\rho \rho}$ (in the same range) would originate an increase
of 1\% in $M_{\text{max}}$, an increase up to 9\% in
$Y_{s}^{\text{max}}$, and a reduction up to 20\%  in $v_s^2(n_\text{max})$. 
Therefore, by including the a $\chi_{\omega\rho}$ term we could have at most $\approx 28\%$
$s$-quarks for $\chi_{\omega\omega}=25$ and $\chi_\rho=0.5$.
However,   the reduction of the speed of sound,
originated from $\chi_{\rho \rho}$, is interesting as it allows to
counterbalence the $\chi_{\omega \omega}$ impact on the speed of sound.  
A CFL phase is favored when the three flavors appear in similar ammounts.  
It could be that more favorable conditions for a CFL phase are still possible with a
correct choice of the couplings $\chi_{\omega\omega}$, that increases the quark branch and speed of
sound, and  $\chi_{\omega\rho}$ that increases the strangeness content and decreases the speed of sound.  
This interplay effect will be explored in a future work.

\subsection{Binary tidal deformability}

The leading tidal parameter of the gravitational-wave signal from a NS merger is the 
effective tidal deformability,
\begin{equation}
\tilde{\Lambda}=\frac{16}{13}\frac{(12q+1)\Lambda_1+(12+q)q^4\Lambda_2}{(1+q)^5},
\label{eq:lambda}
\end{equation}
where $q=M_2/M_1<1$ is the binary mass ratio \cite{Hinderer:2009ca}.
The $\Lambda_{1}\, (M_1)$ and $\Lambda_{2}\, (M_2)$ represent the tidal deformability (mass) of the primary and the secondary NS in the binary, respectively.
The GW170817 event provides 
$\tilde{\Lambda}=300^{+420}_{-230}$ (90\% credible interval), and $0.73\leq q \leq1$ for the binary mass ratio \cite{Abbott:2018wiz}. The chirp mass of the binary system, $M_{\text{chirp}}=(M_1M_2)^{3/5}/(M_1+M_2)^{1/5}$, is measured with a good accuracy during the gravitational wave detection. For the GW170817 event, it was measured to be $1.186^{+0.001}_{-0.001}M_{\odot}$ \cite{Abbott:2018wiz}.\\

\begin{figure}[H]
	\centering
	\includegraphics[width=1.0\columnwidth]{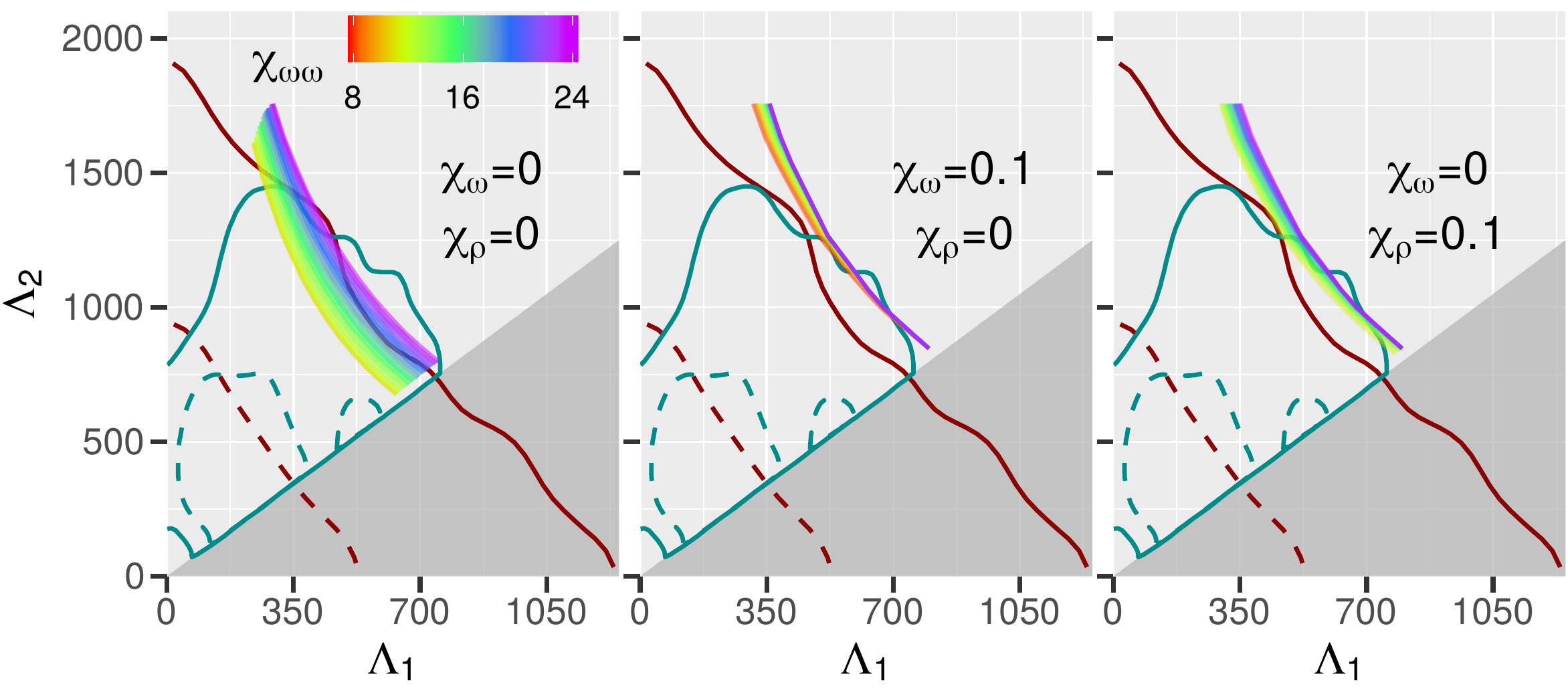}
	\caption{$\Lambda_2-\Lambda_1$ diagrams for binary systems with $M_{\text{chirp}}=1.186M_{\odot}$ and $0.7< q <1$ 
		as a function of $\chi_{\omega\omega}$ (color scale).  
		Three $(\chi_{\omega}, \chi_{\rho})$ sets are presented: $(0,0)$ (left), $(0.1,0)$ (center), and $(0,0.1)$ (right). 
		The	50\% (dashed lines) and 90\% (solid lines)  credible regions
        for the marginalized posteriors $(\Lambda_1,\Lambda_2)$
        from the LIGO/Virgo analysis of the GW170817 event, assuming independent EoS for the NS components
        (brown) \cite{Abbott:2018wiz} and a parametrized EoS assuming
        $M_{\text{max}}\geq1.97 M_{\odot}$ (green) \cite{Abbott:2018exr}. }
\label{fig:lambda1_lambda2}
\end{figure}

In the following, we fix the chirp mass as $M_{\text{chirp}}=1.186M_{\odot}$, and determine  $\Lambda_{1,2}$ for binary systems of mass ratios $0.73\leq q \leq1$. 
The $\Lambda_2-\Lambda_1$ diagrams are shown in Fig. \ref{fig:lambda1_lambda2} as a function of $\chi_{\omega\omega}$ 
(color scale) for three $(\chi_{\omega}, \chi_{\rho})$ sets: $(0,0)$ (left), $(0.1,0)$ (center), and $(0,0.1)$ (right). 
We also show the two distinct credible regions from LIGO/Virgo analysis:
assuming that each NS has a different
EoS, and $\Lambda_{1,2}$ vary independently (brown lines), and using a parametrized EoS with the assumption of $M_{\text{max}}\geq1.97 M_{\odot}$ (dark green lines).
\noindent All hybrid EoS for the $(\chi_{\omega}=0, \chi_{\rho}=0)$ case fall inside the  90\% credible region for mass ratios $0.8\leq q\leq1$. For highly asymmetric binaries, $ 0. 73 \leq q<0.8$, the primary mass $M_1$ can be as large as $1.60M_{\odot}$ while $M_2$ (secondary mass) 
 can be as low as $1.17 M_{\odot}$. The  considerable large radius of $M_1$ for parametrizations with large $\chi_{\rho}$, very close to the purely hadronic EoS result (see solid black line in Fig. \ref{fig:TOV}), results in large $\tilde{\Lambda}$ compared with the LIGO/Virgo analysis (note that we are only considering EoS that reach $1.97M_{\odot}$).
We see that for $(\chi_{\omega}=0, \chi_{\rho}=0)$ all EoS are compatible with LIGO/Virgo 90\% credible region.
A small and finite $\chi_{\omega}$ and/or  $\chi_{\rho}$ reduces the
quark core and thus gives larger values in the $\Lambda_1$-$\Lambda_2$
diagram. The configurations   $(\chi_{\omega},
  \chi_{\rho})=(0.1,0)$ (center) and  $(0,0.1)$ (right) still predict
a large number of hybrid NS compatible with the  90\% credible region of
LIGO/Virgo, for $q \gtrsim 0.85$.

\begin{figure}[!htb]
	\centering
	\includegraphics[width=1.0\columnwidth]{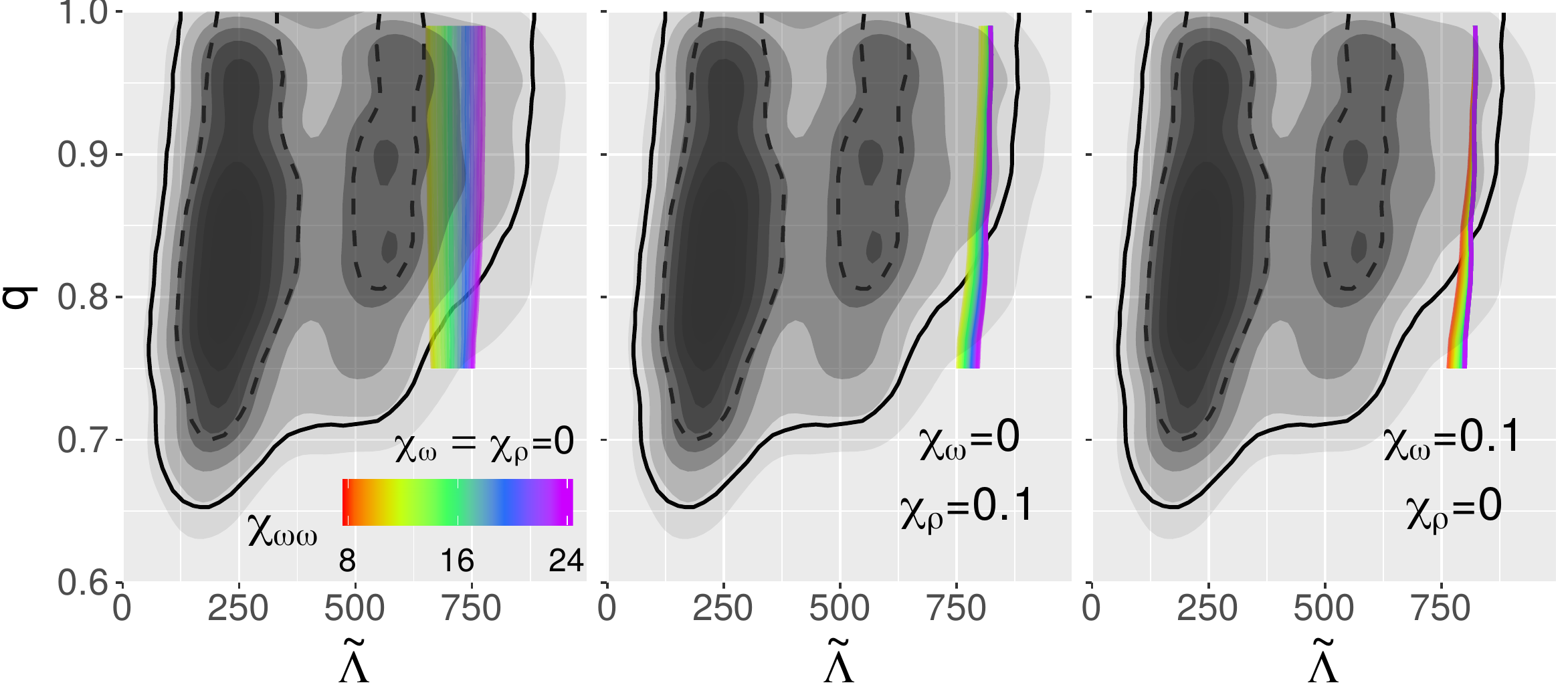}
	\caption{$q-\tilde{\Lambda}$ diagrams for binary systems with $M_{\text{chirp}}=1.186M_{\odot}$ as a function of $\chi_{\omega\omega}$ (color scale).  
		Three $(\chi_{\omega}, \chi_{\rho})$ sets are presented: $(0,0)$ (left), $(0.1,0)$ (center), and $(0,0.1)$ (right). 
	The LIGO/Virgo collaboration results \cite{Abbott:2018wiz} for 
	the probability distribution function of the joint posterior  
	is shown by the gray gradient, and the black dashed and solid 
	lines indicate the 50\% and 90\% credible regions, respectively.}
\label{fig:q_Lambda}
\end{figure}

We compare our set of hybrid EoS and the probability density function (pdf) $P(q,\tilde{\Lambda})$ from the LIGO/Virgo analysis in Fig. \ref{fig:q_Lambda}
(the same sets as in Fig. \ref{fig:lambda1_lambda2}).
Marginalizing the pdf over $q$, i.e., $P(\tilde{\Lambda})=\int P(q,\tilde{\Lambda}) dq$, results in a pdf for $\tilde{\Lambda}$ characterized by $\tilde{\Lambda}=300^{+500}_{-190}$ \cite{Abbott:2018wiz}.
We see that $\tilde{\Lambda}$ depends very weakly on the binary mass ratio $q$. 
Another important observation is that all hybrid EoS shown are within the 90\% credible interval for $q\gtrsim 0.8$ for  $\chi_{\omega}= \chi_{\rho}=0$ and  $q\gtrsim0.85$ if  ($\chi_{\omega},\chi_{\rho})= (0.1,0)$ or (0,0.1). 
The EoS are concentrated around  $\tilde{\Lambda}=650-750$
for $(\chi_{\omega}=0, \chi_{\rho}=0)$, and $\tilde{\Lambda} \sim 800$
for the other two cases. These results show that  all  hybrid stars
studied, both with small or large quark cores which is controlled by the
$\chi_{\omega\omega}$ coupling, are compatible with the GW170817 event.

\section{Conclusions}
\label{conclusions}

We have analyzed the effect of several channel interactions in hadron-quark 
hybrid EoS within the three-flavor NJL model. 
Each hybrid EoS consists of charge-neutral matter in $\beta-$equilibrium 
that has a phase transition from hadronic to quark matter.
We have studied how the stability of hybrid star sequences and their
properties depend on the coupling values. The interaction coupling
considered in this work were: $\chi_{\omega}$, $\chi_{\omega \omega}
$, $\chi_{\rho}$, $\chi_{\rho \rho}$, and $\chi_{\omega \rho}$. Our
analysis showed, however, a weak impact of both $\chi_{\rho \rho}$ and
$\chi_{\omega \rho}$ interactions on the quark EoS, in particular of the first one.
Therefore, we focused our discussion on the $\chi_{\omega}$, $\chi_{\omega \omega} $ and 
$\chi_{\rho}$ channels, and presented a brief estimate of the
  effect of the couplings $\chi_{\rho \rho}$ and
$\chi_{\omega \rho}$. As already seen in \cite{Ferreira:2020kvu}, $\chi_{\omega \omega}$ 
considerably changes the $v_s(n)$ dependence, which turns out to be crucial in supporting massive hybrid NS.  
On the other hand, $\chi_{\rho}$ controls the onset density of the
strange quark, larger values giving lower onset densities and,
consequently, at moderate baryonic densities give rise to a decrease of  the speed of sound.

These effects on the quark EoS are reflected on the properties of
hybrid stars in the following manner: (a) the $\chi_{\omega \omega}$ is
responsible for longer quark branches in the $M(R)$ diagram, larger
central baryonic densities, the onset of quarks in light mass NS and
the possibility of sustaining these quark cores in massive stars with
masses up to $\sim 2.1M_{\odot}$; (b)  the $\chi_{\rho}$ coupling
mainly controls the onset of strangeness in the star,  a earlier onset
occurring with a larger coupling, and its strangeness content, larger
$s$-quark fractions occur for larger couplings. The central
speed of sound decreases as $\chi_{\rho}$ increases.
$\chi_{\rho}$ also influences the
onset of quarks, a larger coupling pushes the onset to larger
densities, decreases the size of the quark branch and gives rise to larger radii. 
Besides, this interaction is responsible for a small
crossing region in the $(M,R)$ diagram taking all the EoS with
different $\chi_{\rho}$ couplings and the other couplings fixed; 
(c)  the $\chi_{\omega}$ is mainly responsible to increasing the 
maximum mass in the $(M,R)$ diagram while pushing the
quark onset to larger densities. This gives rise to shorter quark branches, 
smaller central baryonic densities and smaller strangeness fractions.

Both $ \chi_{\omega \omega}$ and $\chi_{\rho}$ have a competitive
effect on the size of quark cores. While $ \chi_{\omega \omega}$ is
able to generate quark core masses up to $1.53M_{\odot}$, a finite
$\chi_{\rho}$ decreases this effect. However, moderate values
$\chi_{\rho}$ are seen to describe heavier hybrid NS, with smaller
quark cores but with larger strangeness fractions and lower speed of sound
at the NS central densities. In particular, it is possible to have quark
cores with masses of the order of $0.8-1\,M_\odot$ and radii of the
order of 6 km, i.e.,  one third of the star radius.
In other words, $\chi_{\rho}$ is capable of producing massive hybrid
NS, and although with smaller quark cores than taking $\chi_{\rho}=0$
and a larger $ \chi_{\omega \omega}$,  with considerable lower values for $v_s^2$ at the star center.
For any coupling $\chi_{\rho}>0.25$, regardless of the  $\chi_{\omega
\omega}$ value, any quark core in the hybrid stars sequences
contains strange quarks since  the onset density of the strange quarks
happens at lower values than the transition from hadronic to quark
matter.  Under the conditions investigated in the present work, a maximum $s$-quark
fraction of $25\%$ was obtained for a star with a quark core mass
having a third of the total star mass.  Including a $\chi_{\omega\rho}$ term this fraction
could rise up to 28\%.

We have shown that the presently existing mass and tidal
deformability constraints from NS observations
allow for the existence of hybrid stars with a large strangeness
content and
large quark cores. In order to get large $s$-quark content, the
isovector-vector interaction was included. The hybrid stars were built
starting from an hadronic EoS that satisfy presently accepted nuclear
matter properties, DDME2, and considering a quark model that is
constrained to the vacuum properties of several mesons.  Under the conditions considered no twin
stars  \cite{Benic:2014jia} have been found.  In \cite{Benic:2014jia},
twin stars where obtained within a two-flavor NJL model with 4-quark and 8-quark terms, and, 
besides the scalar channels also isoscalar-vector channels were included.

We have compared our set of hybrid stars with the analysis from the GW170817 event. 
A considerable subset of EoS satisfy the 90\% credible region for $\tilde{\Lambda}$. 
The $\chi_{\omega \omega}$ is able to generate massive quark cores,
already present in $1.4M_{\odot}$ stars,  
with  $\tilde{\Lambda}<800$, even for finite but small values of both $\chi_{\rho}$ and $\chi_{\omega}$.
The recently reported GW190814 event of a compact binary coalescence
showed that while the primary component is conclusively a black hole, the secondary component 
with mass of $2.50-2.67M_{\odot}$ remains yet inconclusive \cite{Abbott:2020khf}. 
In the present work, we cannot describe such a massive NS as an hybrid
star described by the present quark model. However, our results are
dependent on the hadronic EoS considered and changing the hadronic part of the hybrid EoS might 
make the EoS hard enough to reach such high mass hybrid stars \cite{Ferreira:2020evu}. 
In this case, the quark content of such stars will be very small and the tidal deformability 
of stars with low masses, within such models will be given by the hadronic sector of the EoS.

\begin{acknowledgments}

This work was partially supported by national funds from FCT (Fundação para a Ciência e a Tecnologia, I.P, Portugal) under the IDPASC Ph.D. program (International Doctorate Network in Particle Physics, Astrophysics and Cosmology), with the Grant No. PD/\-BD/128234/\-2016 (R.C.P.), under the Projects No. UID/\-FIS/\-04564/\-2019, No. UID/\-04564/\-2020, and No. POCI-01-0145-FEDER-029912 with financial support from Science, Technology and Innovation, 
in its FEDER component, and by the FCT/MCTES budget through national funds (OE).
\end{acknowledgments}

\onecolumngrid
\appendix

\section{Hybrid stars properties}
We summarize some properties of the
hybrid NS obtained with  $\chi_{\omega}=0$ and
$\chi_{\omega}=0.1$ in Tables \ref{comgea0} and \ref{comgea1}, respectively,
for several values of  $\chi_{\omega\omega}$ and $\chi_{\rho}$.

\begin{table*}[!htb]
\centering
\begin{tabular}{cc|ccccccccccc}
  \hline
  \hline
&  & $M_{\text{QC}}$ & $R_{\text{QC}}$ & $M_{\text{max}}$ & $R_{\text{max}}$ & $n_{q}$ & $M_q$  & $n_{\text{max}}$ & $Y^{\text{max}}_s$ & $v_s^2(n_{\text{max}})$ & $R_{1.4M_{\odot}}$ & $\Lambda_{1.4M_{\odot}}$\\
$\chi_{\rho}$& $\chi_{\omega\omega}$ & [$M_{\odot}$] & [km] & [$M_{\odot}$] & [km] & [fm$^{-3}$] & [$M_{\odot}$] & [fm$^{-3}$] &  & [c$^2$] & [km] & \\
\hline
 $0.0 $ & $0$   & $1.21$ & $8.69$ & $1.71$ &  $11.69$&  $0.33$ & $0.92$ & $0.95$ & $0.17$ & $0.32$ & $12.54$ & $452$\\
 $0.0 $ & $5$   & $1.41$ & $8.77$ & $1.84$ &  $11.31$&  $0.33$ & $0.98$ & $1.03$ & $0.20$ & $0.43$ & $12.68$ & $493$\\
 $0.0 $ & $10$   & 1.57 & 8.82 & 1.96 &  11.03 &  0.34 & 1.04 & 1.07 & 0.20 & 0.66 & 12.81 & 538\\
 $0.0 $ & $15$  & $1.68$ & $8.89$ & $2.05$  &  $10.96$&  $0.34$ & $1.10$ & $1.06$ & $0.20$ & $0.83$ & $12.91$ & $578$\\
 $0.0$ & $20$  & 1.74 & 8.96 & 2.13 & 11.01 &  0.35 & 1.18 & 1.03 & 0.2 & 0.93 & 13.02 & 621\\
 $0.2$ & $0$   & 0.76 & 7.24 & 1.72 & 12.26 &  0.39 & 1.38 & 0.84 & 0.18 & 0.31 & 13.19 & 690\\
 $0.2$ & $5$   & 0.97 & 7.55 & 1.84 & 11.87 &  0.40 & 1.46 & 0.93 & 0.21 & 0.37 & 13.20 & 705\\
 $0.2$ & $10$  & 1.20 & 7.78 & 1.95 & 11.43  &  0.41 & 1.53 & 1.00 & 0.22 & 0.61 & 13.20 & 705\\
 $0.2$ & $15$  & 1.32 & 7.89 & 2.04 & 11.25 &  0.42 & 1.63 & 1.02 & 0.23 & 0.78 & 13.20 &  705\\
 $0.2$ & $20$  & 1.37 & 7.92 & 2.12 & 11.26 &  0.43 & 1.71 & 1.00 & 0.22 & 0.89 & 13.20 &  705\\
 $0.4$ & $0$   & 0.30 & 5.09 & 1.77 & 12.82  &  0.48 & 1.71 & 0.72 & 0.17 & 0.29 & 13.20 & 705\\
 $0.4$ & $5$   & 0.47 & 5.78 & 1.86 & 12.52 & 0.49 & 1.77 & 0.79 & 0.20 & 0.31 &  13.20 &  705\\
 $0.4$ & $10$  & 0.75 & 6.52 & 1.95 & 12.02 &  0.50 & 1.84 & 0.89 & 0.22 & 0.50 &  13.20 & 705\\
 $0.4$ & $15$  & 0.95 & 6.88 & 2.05 & 11.68 &  0.51 & 1.91 & 0.94 & 0.23 & 0.69 & 13.20 &  705\\
 $0.4$ & $20$  & 1.03 & 7.01 & 2.13 & 11.59 &  0.52 & 1.98 & 0.94 & 0.23 & 0.82 & 13.20 &  705\\
   \hline
     \hline
\end{tabular}
	\caption{Several NS properties for each different hybrid EoS
          $(\chi_{\rho},\chi_{\omega\omega},\chi_{\omega}=0)$: quark
          core ($M_{\text{QC}}$), quark radii ($R_{\text{QC}}$), mass
          of the heaviest star ($M_{\text{max}}$), radii of the
          heaviest star ($R_{\text{max}}$), onset density of quarks
          ($n_q$), lightest NS mass with quark content ($M_q$),
          central density of the heaviest star ($n_{\text{max}}$),
          strange quark fraction at $n_{\text{max}}$
          ($Y_s^{\text{max}}$), squared speed of sound at
          $n_{\text{max}}$ ($v_s^2(n_{\text{max}})$), radii
          ($R_{1.4M_{\odot}}$) and tidal deformability
          ($\Lambda_{1.4M_{\odot}}$) of a $1.4M_{\odot}$ NS.}
        \label{comgea0}
\end{table*}

\begin{table*}[!htb]
\centering
\begin{tabular}{cc|ccccccccccc}
  \hline
    \hline
&  & $M_{\text{QC}}$ & $R_{\text{QC}}$ & $M_{\text{max}}$ & $R_{\text{max}}$ & $n_{q}$ & $M_q$  & $n_{\text{max}}$ & $Y^{\text{max}}_s$ & $v_s^2(n_{\text{max}})$ & $R_{1.4M_{\odot}}$ & $\Lambda_{1.4M_{\odot}}$\\
$\chi_{\rho}$& $\chi_{\omega\omega}$ & [$M_{\odot}$] & [km] & [$M_{\odot}$] & [km] & [fm$^{-3}$] & [$M_{\odot}$] & [fm$^{-3}$] &  & [c$^2$] & [km] & \\
\hline
 $0.0 $ & $0$   & $1.03 $& $7.99$ & $1.80$ &  $11.99$ &  $0.37$ & $1.27$  & $0.90$ & $0.16$ & $0.33$ & $13.07$ &  $642$ \\
 $0.0 $ & $5$   & $1.21 $& $8.20$ & $1.91$ &  $11.71$ &  $0.38$ & $1.34$  & $0.95$ & $0.17$ & $0.40$ & $13.16$ &  $684$ \\
 $0.0$ & $10$   & 1.35 & 8.29 & 2.01 &  11.48 &  0.39 & 1.42  & 0.98 & 0.18 & 0.58 & 13.20 &  707 \\
 $0.0 $ & $15$  & $1.44 $& $8.35$ & $2.10$ &  $11.39$ &  $0.40$ & $1.50$  & $0.98$ & $0.18$ & $0.73$ & $13.20$ &  $710$ \\
 $0.0$ & $20$   &  1.49 & 8.37 & 2.17 &  11.39 &  0.41 & 1.59  & 0.97 & 0.18 & 0.84 & 13.20 &  705 \\
 $0.2$ & $0$   & 0.51 & 6.15 & 1.84 &  12.59 &  0.45 & 1.69 & 0.78 & 0.16 & 0.32 & 13.20 &  705 \\
 $0.2$ & $5$   & 0.66 & 6.53 & 1.94 &  12.35 &  0.46 & 1.76  & 0.83 & 0.18 & 0.34 & 13.20 &  705 \\
 $0.2$ & $10$   & 0.82 & 6.82 & 2.02 &  12.08 &  0.47 & 1.84  & 0.87 & 0.19 & 0.49 & 13.20 &  705 \\
 $0.2$ & $15$   & 0.93 & 6.98 & 2.11 &  11.9 &  0.49 & 1.92  & 0.9 & 0.2 & 0.64 & 13.20 &  705 \\
 $0.2$ & $20$   & 0.98 & 7.00 & 2.18 &  11.83 &  0.51 & 2.0 & 0.9 &  0.2 & 0.75 & 13.20 &  705 \\
 $0.4$ & $0$   & 0.14 & 3.84 & 1.91 &  13.07 &  0.54 & 1.88  & 0.67 & 0.16 & 0.3 & 13.20 &  705 \\
 $0.4$ & $5$   & 0.22 & 4.47 & 1.98 &  12.92 &  0.55 & 1.95  & 0.71 & 0.17 & 0.31 & 13.20 &  705 \\
 $0.4$ & $10$   & 0.35 & 5.11 & 2.06 &  12.7 &  0.56 & 2.01  & 0.75 & 0.19 & 0.39 & 13.20 &  705 \\
 $0.4$ & $15$   & 0.48 & 5.57 & 2.13 &  12.48 &  0.57 & 2.08 & 0.79 & 0.2 & 0.52 & 13.20 &  705 \\
 $0.4$ & $20$   & 0.57 & 5.81 & 2.2 &  12.32 &  0.58 & 2.14  & 0.81 & 0.2 & 0.64 & 13.20 &  705 \\
   \hline
     \hline
\end{tabular}
\caption{Several NS properties for each different hybrid EoS
  $(\chi_{\rho},\chi_{\omega\omega},\chi_{\omega}=0.1)$: quark core
  ($M_{\text{QC}}$), quark radii ($R_{\text{QC}}$), mass of the
  heaviest star ($M_{\text{max}}$), radii of the heaviest star
  ($R_{\text{max}}$), onset density of quarks ($n_q$), lightest NS
  mass with quark content ($M_q$), central density of the heaviest
  star ($n_{\text{max}}$), strange quark fraction at $n_{\text{max}}$
  ($Y_s^{\text{max}}$), squared speed of sound at $n_{\text{max}}$
  ($v_s^2(n_{\text{max}})$), radii ($R_{1.4M_{\odot}}$) and tidal
  deformability ($\Lambda_{1.4M_{\odot}}$) of a $1.4M_{\odot}$ NS.}
        \label{comgea1}
\end{table*}
\normalem
\newpage
\twocolumngrid

%

\end{document}